\documentclass[twocolumn,amsmath,showpacs,nofootinbib]{revtex4-1}
\usepackage{graphicx}
\usepackage{dcolumn}
\usepackage{bm}
\usepackage{color} 
\usepackage{slashed}
\begin{document}
\newcommand{\hs}{\hspace*{0.5cm}}
\newcommand{\vs}{\vspace*{0.5cm}}
\newcommand{\be}{\begin{equation}}
\newcommand{\ee}{\end{equation}}
\newcommand{\bea}{\begin{eqnarray}}
\newcommand{\eea}{\end{eqnarray}}
\newcommand{\ben}{\begin{enumerate}}
\newcommand{\een}{\end{enumerate}}
\newcommand{\bde}{\begin{widetext}}
\newcommand{\ede}{\end{widetext}}
\newcommand{\nn}{\nonumber}
\newcommand{\crn}{\nonumber \\}
\newcommand{\Tr}{\mathrm{Tr}}
\newcommand{\non}{\nonumber}
\newcommand{\noi}{\noindent}
\newcommand{\al}{\alpha}
\newcommand{\la}{\lambda}
\newcommand{\bet}{\beta}
\newcommand{\ga}{\gamma}
\newcommand{\va}{\varphi}
\newcommand{\om}{\omega}
\newcommand{\pa}{\partial}
\newcommand{\+}{\dagger}
\newcommand{\fr}{\frac}
\newcommand{\bc}{\begin{center}}
\newcommand{\ec}{\end{center}}
\newcommand{\Ga}{\Gamma}
\newcommand{\de}{\delta}
\newcommand{\De}{\Delta}
\newcommand{\ep}{\epsilon}
\newcommand{\varep}{\varepsilon}
\newcommand{\ka}{\kappa}
\newcommand{\La}{\Lambda}
\newcommand{\si}{\sigma}
\newcommand{\Si}{\Sigma}
\newcommand{\ta}{\tau}
\newcommand{\up}{\upsilon}
\newcommand{\Up}{\Upsilon}
\newcommand{\ze}{\zeta}
\newcommand{\ps}{\psi}
\newcommand{\Ps}{\Psi}
\newcommand{\ph}{\phi}
\newcommand{\vph}{\varphi}
\newcommand{\Ph}{\Phi}
\newcommand{\Om}{\Omega}
\newcommand{\AdrHEPC}{Phenikaa Institute for Advanced Study and Faculty of Basic Science, Phenikaa University, Yen Nghia, Ha Dong, Hanoi 100000, Vietnam}

\title{Flipping principle for neutrino mass and dark matter} 

\author{Phung Van Dong} 
\email{dong.phungvan@phenikaa-uni.edu.vn}
\affiliation{\AdrHEPC} 
\date{\today}

\begin{abstract}
Flipping a symmetry often leads to a more fundamental symmetry and new physics insight. Applying this principle to the standard model electroweak symmetry, we obtain a novel gauge symmetry, which defines dark charge besides electric charge, neutrino mass mechanism, and resultant dark parity as residual flipped symmetry. The dark parity divides the model particles into two classes: odd and even. The dark matter candidate transforms as a fermion or a scalar singlet, having a mass below the electron mass, being stabilized by the dark parity. Scenarios for consistent neutrino mass generation and dark matter relic are proposed. The nature and further implication of the flipping approach are determined.
\end{abstract}


\maketitle

The idea of flipping a symmetry leading to a relevant new physics is not new, but its principle has not explicitly been understood. Further, the implication of flipped symmetry has not significantly been studied.   

Take, for instance, the $SU(2)_L$ symmetry of weak isospin $T_{1,2,3}$ in the V--A theory \cite{fg,sm,s}. The matter content transforms as isodoublets $l_L=(\nu_L\ e_L)$ and $q_L=(u_L\ d_L)$.\footnote{Although generation indices are suppressed, three generations of fermions are always taken into account.} The key for flipping $SU(2)_L$ lies at the electric charge ($Q$). Indeed, the electric charge of multiplets is determined as $Q=\mathrm{diag}(0,-1)$ for $l_L$ and $Q=\mathrm{diag}(2/3,-1/3)$ for $q_L$. This yields $[Q,T_1\pm i T_2]=\pm(T_1\pm i T_2)\neq 0$ and $\Tr Q\neq 0$. Thus, $Q$ neither commutes nor closes algebraically with $T_{1,2,3}$.\footnote{Otherwise algebraic closure requires $Q$ to be some generator of $SU(2)_L$, i.e. $Q=x_i T_i$, which leads to a contradiction $\Tr Q=0$.} The nonclosure requires $SU(2)_L$ flipped (i.e. enlarged) to $SU(2)_L\otimes U(1)_Y$ by symmetry principles, called gauge completion, where $Y\equiv Q-T_3$ is a new Abelian charge, well-known as hypercharge. A new observation is that since $T_3$ is gauged, $Q$ and $Y$ must simultaneously be gauged as a result of the noncommutation. Hence, the flip leads to electroweak theory \cite{g,w,as}, predicting the neutral current and the need of right-handed fermion singlets $e_R$, $u_R$, and $d_R$ for canceling the $U(1)_Y$ anomalies.  

Now a curious question is that can neutrino mass and dark matter be flipped from $SU(2)_L$ too? Since the electric charge is theoretically not fixed \cite{ecq,ecq1,ecq2,ecq3}, we suppose a variant of it, called dark charge ($D$), such that $D=\mathrm{diag}(1,0)$ for $l_L$ and $D=\mathrm{diag}(1/3,-2/3)$ for $q_L$. The dark charge for the Higgs doublet $\phi=(\phi^+\ \phi^0)$ is $D=\mathrm{diag}(1,0)$, which is assigned so that $D$ is not broken by the electroweak vacuum. Analogous to $Q$, the dark charge $D$ neither commutes nor closes algebraically with $SU(2)_L$, which yields, by the flipping principle, a novel gauge symmetry $SU(2)_L\otimes U(1)_{N}$, where $N\equiv D-T_3$. Because the charges $Y$ and $N$ are linearly independent, the full gauge symmetry takes the form, 
\be SU(2)_L\otimes U(1)_Y\otimes U(1)_{N},\label{gs}\ee apart from the QCD group. Besides the mentioned fermion singlets, the right-handed neutrino $\nu_R$ is now required to cancel the $[\mathrm{Gravity}]^2U(1)_{N}$ anomaly. It is checked that all the other anomalies vanish too.  
For comparison, the $Q$ and $D$ charges are collected in Table~\ref{tab1}, while the $SU(2)_L$ representations and $Y$, $N$ charges are supplied in Table \ref{tab2}. Here, the singlet scalar $\chi$ is necessarily included to break $U(1)_{N}$ symmetry and generate right-handed neutrino mass.\footnote{A general version of dark charge can be constructed in terms of arbitrary $\delta$ parameter, in which $D(\nu)=\delta$, $D(e)=\delta-1$, $D(u)=2/3-\delta/3$, and $D(d)=-1/3-\delta/3$, for every both left-handed and right-handed fermion components $f_{L,R}$, while $D(\chi)=-2\delta$ and the dark charge of $\phi$ is retained. However, the present case with $\delta=1$ makes dark matter stability easily.}  
\begin{table}[h]
\bc
\begin{tabular}{c|ccccccc}
\hline\hline
Field & $\nu$ & $e$ & $u$ & $d$ & $\phi^+$ & $\phi^0$ & $\chi$ \\
\hline
$Q$ & 0 & $-1$ & $2/3$ & $-1/3$ & 1 & 0 & 0
\\ \hline
$D$ & 1 & 0 & $1/3$ & $-2/3$ & 1 & 0 & $-2$\\
\hline\hline
\end{tabular}
\caption[]{\label{tab1} $Q$ and $D$ charges of the model particles.}
\ec
\end{table}          
\begin{table}[h]
\bc
\begin{tabular}{c|cccccccc}
\hline\hline
Multiplet & $l_L$ & $q_L$ & $\nu_R$ & $e_R$ & $u_R$ & $d_R$ & $\phi$ & $\chi$ \\
\hline 
$SU(2)_L$ & 2 & 2 & 1 & 1& 1& 1& 2 & 1 \\ \hline
$Y$ & $-1/2$ & $1/6$ & 0 & $-1$ & $2/3$ & $-1/3$ & $1/2$ & 0 \\ \hline
$N$ & $1/2$ & $-1/6$ & $1$ & $0$ & $1/3$ & $-2/3$ & $1/2$ & $-2$\\
\hline\hline
\end{tabular}
\caption[]{\label{tab2} $SU(2)_L$, $Y$, and $N$ quantum numbers of the model multiplets.}
\ec
\end{table} 

The $U(1)_{N}$ symmetry must be broken at a high energy scale for the model consistency. This can be done when the scalars develop vacuum expectation values (VEVs), $\langle \chi\rangle = \La/\sqrt{2}$ and $\langle \phi \rangle = (0\ v)/\sqrt{2}$, such that $\La\gg v =246$ GeV. Thus, the scheme of gauge symmetry breaking is \bc \begin{tabular}{c} $SU(2)_L\otimes U(1)_Y\otimes U(1)_{N}$ \\
$\downarrow\La$\\
$SU(2)_L\otimes U(1)_Y\otimes D'_P$\\
$\downarrow v$\\
$U(1)_Q\otimes D_P$ \end{tabular}\ec
where the first step implies a dark parity $D'_P$ as a residual symmetry of $U(1)_N$, while the second step determines the electroweak symmetry breaking, as usual, and the resultant dark parity $D_P$, which is a residual symmetry of $SU(2)_L\otimes D'_P$ or $D=T_3+N$.

Indeed, since the $N$ charge of $\chi$ is $N(\chi)=-2\neq 0$, the VEV of $\chi$ breaks $N$, i.e. $N\langle \chi\rangle =-\sqrt{2}\La\neq 0$, which does not annihilate the $\chi$ vacuum. The $U(1)_N$ symmetry is not completely broken, since it may conserve a residual symmetry $D'_P=e^{i\al N}$, such that $D'_P \langle \chi\rangle = \langle \chi\rangle$. Hence, $e^{i\al (-2)}=1$, or equivalently, $\al=k\pi$ for $k$ integer. The residual symmetry takes the form, \be D'_P=e^{i k \pi N}=(-1)^{kN}.\ee Next, $\phi$ has charges $T_i=\fr 1 2 \sigma_i \neq 0$, $Y=\fr 1 2\neq 0$, and $N=\fr 1 2 \neq 0$. The VEV of $\phi$, $\langle \phi \rangle =(0\ v)/\sqrt{2}$ breaks all these charges, since $T_i\langle \phi\rangle \neq 0$, $Y\langle \phi\rangle = (0\ v)/2\sqrt{2} \neq 0$, and $N\langle \phi\rangle = (0\ v)/2\sqrt{2} \neq 0$, which do not annihilate the electroweak vacuum. A residual charge of $SU(2)_L\otimes U(1)_Y\otimes D'_P$ must be composed of $R\equiv a_i T_i + b Y + c N$, and $R$ annihilates the electroweak vacuum, $R\langle \phi\rangle =0$. It follows that $a_1=a_2=0$ and $a_3=b+c$, hence \be R=b(T_3+Y)+c(T_3+N)=b Q+ c D.\ee Here, the charge combinations $Q=T_3+Y$ and $D=T_3+N$ commute, $[Q,D]=0$, and separately conserve the electroweak vacuum, since $Q\langle \phi\rangle =0 = D \langle \phi\rangle$. The former charge combination defines the electromagnetic symmetry $U(1)_Q$, as usual, which annihilates both $v,\La$ and is the residual symmetry of $SU(2)_L\otimes U(1)_Y$. Whereas, the latter charge combination defines a final dark parity $D_P=e^{i\beta D}$, which transforms a general field as $\Phi\rightarrow \Phi'=D_P\Phi$. As obtained, $D$ always annihilates the $\phi$ vacuum, but similarly to $D'_P$, it must conserve the $\chi$ vacuum, i.e. $D_P \langle \chi\rangle =\langle \chi\rangle$, which matches $\beta=\al=k\pi$ for $k=0,\pm1,\pm2,\cdots$. The dark parity is \be D_P=e^{i k\pi D}=(-1)^{kD},\label{dpa1}\ee
which is a residual symmetry of $SU(2)_L\otimes D'_P$. 

Three remarks are in order
\ben 
\item A dark parity similar to $D'_P$ can be recognized in the $U(1)_{B-L}$ theory or other models with an extra $U(1)$ factor. 
\item The dark parity $D_P$ is newly achieved, which is shifted from $D'_P$ by the electroweak symmetry breaking, a consequence of the noncommutative dark charge existence proposed by this work. 
\item The difference between $D_P$ and $D'_P$ is that $D'_P$ commutes with the electroweak symmetry and transforms the same for every component particle in gauge multiplet, whereas $D_P$ does not commute with the electroweak symmetry and transforms differently for component particles in gauge multiplet that have different $T_3$ values, respectively. Hence, $D_P$ provides a possibility for unifying normal matter and dark matter in gauge multiplet,\footnote{Comparing to supersymmetry, normal matter (particle) and dark matter (sparticle) are unified in supermultiplet.} as well as suppressing unwanted VEV directions (or components) from turning on.      
\een         

With the dark charge values in Table \ref{tab1}, we obtain $D_P=1$ for a minimal value of $k=6$. Thus, the residual symmetry is homomorphic to $Z_6=\{1,p,p^2,p^3,p^4,p^5\}$, where $p\equiv (-1)^D$ and $p^6=1$. This group contains an invariant (or normal) subgroup $Z_2=\{1,p^3\}$ and we can factorize $Z_6=Z_2\otimes Z_3$, where $Z_3=Z_6/Z_2=\{Z_2,\{p,p^4\},\{p^2,p^5\}\}$ is the quotient group of $Z_6$ by $Z_2$. Thus, the theory automatically conserves both the residual symmetries, $Z_2$ and $Z_3$, which can contain the scenarios of multicomponent dark matter, where stable candidates transform nontrivially under $Z_2$ and $Z_3$, respectively. A detailed study of multicomponent dark matter is out of scope of this work, to be published elsewhere.   

Consider only the residual symmetry $Z_2$ which is generated by $p^3$.\footnote{We can build a theory that the residual symmetry coincides with the invariant subgroup $Z_2$ by supposing an extra heavy scalar to have $N=2/3$ for $N,D$ breaking and then integrating this scalar out. But, it is not necessary since the present theory always conserves the quotient group $Z_3$, which is not further discussed.} We redefine the dark parity to be $D_P=p^3=(-1)^{3D}$, which corresponds to (\ref{dpa1}) for $k=\pm3$. The dark parity is conveniently rewritten as \be D_P=(-1)^{3(T_3+N)+2s}\ee after substituting $D=T_3+N$ and multiplying the spin parity $(-1)^{2s}$, which is conserved by the Lorentz symmetry. The dark parity of particles is summarized in Table~\ref{tb3}, where $Z'$ is the $U(1)_{N}$ gauge boson. Note that $\phi^\pm$ is a Goldstone boson eaten by $W^\pm$.  
\begin{table}
\bc
\begin{tabular}{l|cccccccccccc}
\hline\hline
Particle & $\nu$ & $e$ & $u$ & $d$ & $\phi^+$ & $\phi^0$ & $\chi$ & $W$ & $\ga$ & $Z$ & $Z'$\\
\hline
$D_P$ & $1$ & $-1$ & $1$ & $-1$ & $-1$ & $1$ &  1 & $-1$ & 1 & 1 & 1 \\
\hline\hline
\end{tabular}
\caption[]{\label{tb3} Dark parity of the model particles.}
\ec
\end{table}

From Table \ref{tb3}, the electron is the lightest physical particle that is $D_P$ odd. We assume two candidates, a vector-like fermion ($n$) and a scalar ($\eta$), transforming under the gauge symmetry in equation (\ref{gs}) as \be n\sim (1,0,2r),\hs \eta\sim (1,0,2r-1),\ee for $r$ integer, which all couple to $\nu_R$ through $y\bar{n}_L \eta\nu_R$.\footnote{Especially, when $r=0$ we need only introduce the left chiral component $n_L$ (i.e. $n_R$ is omitted), since it does not contribute to the anomalies.} The fields $n,\eta$ are all $D_P$ odd,
\be D_P(n)=-1=D_P(\eta).\ee They possess a mass $\mathcal{L}\supset -m_n\bar{n}n-m^2_\eta \eta^*\eta$, where $m^2_\eta>0$ may be shifted by an amount after symmetry breaking, since $\eta$ couples to other Higgs fields. One of them should be lighter than $e$, responsible for dark matter.\footnote{Hence, the dark matter candidate is stabilized by dark parity conservation, while the electron stability is always ensured by electric charge conservation, by contrast.}  

The model reveals a seesaw mechanism for neutrino mass generation. Indeed, the scalar singlet $\chi$ couples to $\nu_R$ through $\fr 1 2 f^\nu \bar{\nu}^c_R\chi \nu_R$, yielding a large Majorana mass $m_R=-f^\nu \La/\sqrt{2}$ for $\nu_R$. Additionally, $\phi$ couples to both $l_L$ and $\nu_R$ through $h^\nu \bar{l}_L\tilde{\phi} \nu_R$, leading to a Dirac neutrino mass $m_D=-h^\nu v/\sqrt{2}$. Applying the seesaw formula for $\La\gg v$, this provides observed neutrino ($\sim \nu_L$) mass,
\be m_\nu \simeq - m_D m^{-1}_R m_D^T = h^\nu (f^\nu)^{-1}(h^\nu)^T \fr{v^2}{\sqrt{2}\La}.\label{nm} \ee The heavy Majorana neutrino $(\sim \nu_R)$ obtains a mass $\sim m_R$, retained at $\La$ scale. 

{\it Large scale seesaw scenario}. The predicted mass in~(\ref{nm}) coincides with the observation, $m_\nu\sim 0.1$ eV, if 
\be \La\sim [(h^\nu)^2/f^\nu]\times 10^{14}\ \mathrm{GeV}\sim 10^{14}\ \mathrm{GeV},\ee
which is proportional to the inflation scale driven by the $U(1)_N$ dynamics, where we assume $(h^\nu)^2/f^\nu\sim 1$ \cite{dong,dong1}. 

This scenario can explain the baryon asymmetry via standard leptogenesis set by CP-violating decay of $\nu_{R}$ to normal matter $e\phi^+$ \cite{fy}. 

\begin{figure}[h]
\bc
\includegraphics[scale=0.8]{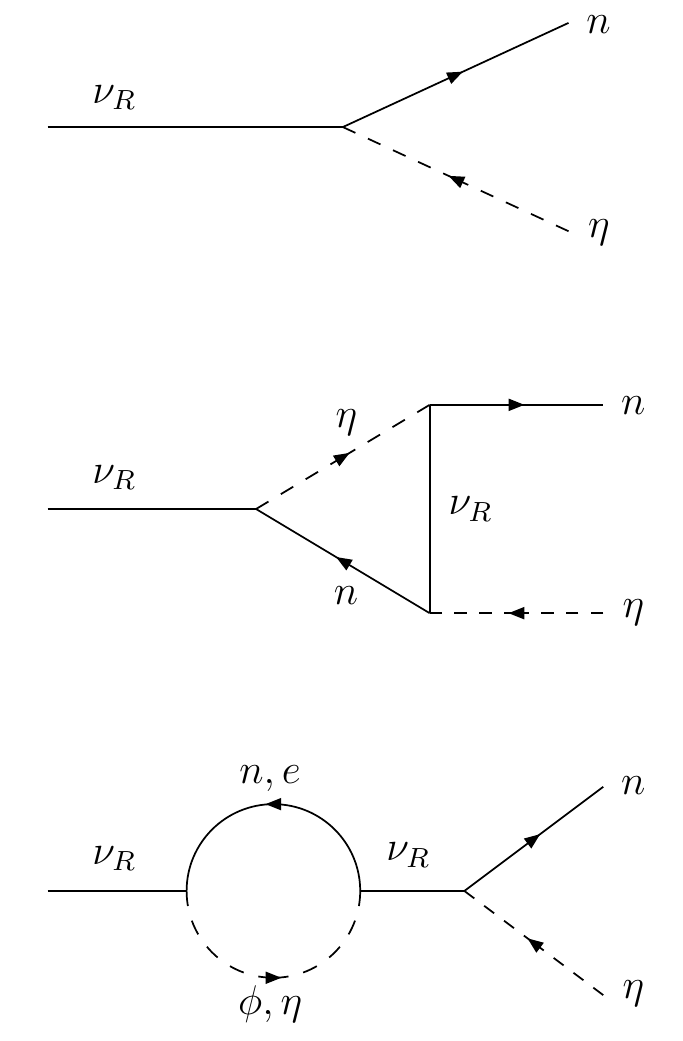}
\caption{\label{fig1} CP-violating decay of $\nu_R$ that produces asymmetric dark matter.} 
\ec
\end{figure}

In the same manner, $\nu_{R}$ can decay CP-asymmetrically to a pair of dark matter $n \eta^*$ through the $y$ coupling, explicitly written as $y_j\bar{n}_L \eta\nu_{jR}$ for $j=1,2,3$. The process is presented by Feynman diagrams in Fig \ref{fig1}. 
Consider the heavy Majorana mass matrix to be flavor diagonal, $m_R=\mathrm{diag}(m_{\nu_{1R}},m_{\nu_{2R}},m_{\nu_{3R}})$, and hierarchical, $m_{\nu_{1R}}\ll m_{\nu_{2,3R}}$.
The CP asymmetry parameter is evaluated as \bea \epsilon_{\mathrm{DM}}&=&\fr{\Ga(\nu_{1R}\rightarrow n\eta^*)-\Ga(\nu_{1R}\rightarrow \bar{n}\bar{\eta}^*)}{\Ga(\nu_{1R}\rightarrow n\eta^*)+\Ga(\nu_{1R}\rightarrow \bar{n}\bar{\eta}^*)}\crn
&\simeq& -\fr{3}{16\pi y^*_1y_1}\sum_{j\neq 1}\Im\left[(y^*_j y_1)^2\right]\fr{m_{\nu_{1R}}}{m_{\nu_{jR}}}. \eea

The observation $\Om_{\mathrm{DM}}\simeq 5 \Om_B$ leads to $m_{\mathrm{DM}}/m_p\simeq 5\eta_B/\eta_{\mathrm{DM}}$, where $\eta_B=-(28/79)\eta_L$ is related to the usual CP asymmetry $\epsilon_L$, governed by $\nu_{1R}\rightarrow e\phi$ decay, and $m_p$ denotes the proton mass. Both thermal and nonthermal leptogenesises imply $\eta_L/\eta_{\mathrm{DM}}\sim \epsilon_L/\epsilon_{\mathrm{DM}}$.\footnote{In nonthermal leptogenesis, the right-handed neutrino produced by inflaton decay $\chi\rightarrow \nu_R\nu_R$, the equality happens.} It follows that $m_{\mathrm{DM}}/m_p\sim -\epsilon_L/\epsilon_{\mathrm{DM}}$. Assuming $m_{\nu_{2,3R}}=10^{2}m_{\nu_{1R}}$, $\epsilon_L\sim 10^{-7}$, and $y_{2,3}=e^{-i\theta}y_1$ with real $y_1$, this yields 
\be m_{\mathrm{DM}}\sim \fr{10^{-4}m_p}{y^2_1\sin(2\theta)}.\ee Taking $y_1\sim 1$ and $\theta\sim \pi/4$, the model predicts (fermion or scalar) dark matter mass to be $m_{\mathrm{DM}}\sim 0.1$ MeV.    
 
{\it Low scale seesaw scenario}. The predicted mass~(\ref{nm}) yields a TeV seesaw scale, i.e. $\La\sim $ TeV, if $(h^\nu)^2/f^\nu$ is suitably small, e.g. $f^\nu\sim 0.1$ and $h^\nu\sim 10^{-6}$ similar to electron Yukawa coupling. 

In this case $Z'$ and $\chi$ can be reached by particle colliders. The LEPII looks for $Z'$ boson via channel $e^+e^-\rightarrow \mu^+\mu^-$, constraining its mass and coupling to be $m_{Z'}/(g_N/2)>6\ \mathrm{TeV}$ \cite{lep}. Since $m_{Z'}\simeq 2g_N \La$, it follows that $\La>1.5$ TeV. The LHC \cite{lhc} searches for dilepton signals mediated by $Z'$, obtaining a mass bound 4 TeV for $Z'$ coupling similar to $Z$, which translates to a bound for $\La$ analogous to the LEPII search. Since $\phi,\chi$ slightly mix, diphoton signals mediated by $\chi$ are appropriately suppressed, in agreement with \cite{lhc1,lhc2}. In other words, $\chi$ may pick up a mass at TeV.         

The dark matter relic can be produced through a freeze-in mechanism \cite{freezein} with narrow decay $\nu_{1R}\rightarrow n\eta^*$, where $\nu_{1R}$ and one of the two candidates (either $n$ or $\eta$) are in thermal equilibrium with standard model plasma. Note that the right-handed neutrino and the dark field in thermal equilibrium are always maintained by their gauge interaction with $Z'$, i.e. $\mathrm{SM}+ \mathrm{SM}\leftrightarrow Z'\leftrightarrow \nu_R\nu_R (\mathrm{either}\ n \bar{n}\ \mathrm{or}\ \eta\eta^*)$, respectively. 

The decay rate is $\Ga_{\nu_{1R}}=(y^2_1/32\pi)m_{\nu_{1R}}$. Generalizing the result in \cite{freezein}, the relic density is
\be \Om_{\mathrm{DM}}h^2\sim 0.1\left(\fr{y_1}{10^{-8}}\right)^2 \left(\fr{300\ \mathrm{GeV}}{m_{\nu_{1R}}}\right)\left(\fr{m_{\mathrm{DM}}}{0.1\ \mathrm{MeV}}\right),\ee implying dark matter mass at $0.1$ MeV for the correct density, $y_1\sim 10^{-8}$, and the lightest right-handed neutrino with a $300$ GeV mass.

{\it Nature and further implication of the flipping principle}. The crucial difference of the flipping approach from other  approaches based upon an extra $U(1)$ factor, e.g. $U(1)_N$, is the proposal of a noncommutative dark charge. Indeed, the $U(1)_N$ factor arises to be a consequence of the algebraic closure of dark charge and that the local symmetry nature of dark charge and the resultant dark parity result from its noncommutation with the electroweak symmetry. In contrast, the gauging of the extra $U(1)$ factor as a starting point is not required on the theoretical ground and eventually leads to a commutative dark parity like the matter parity or a $Z_2$. That said, the flipping principle overhauls the current theories of neutrino mass and dark matter. Take, for instance,
\ben
\item Scotogenic theory \cite{ema}: Add a second Higgs doublet \be \phi'=(\phi'^+\ \phi'^0)\sim (2,1/2,-1/2)\ee to our present model which couples $\bar{l}_L \tilde{\phi}' n^c_L$, where we assume $r=0$ and $n_L$ possessing a Majorana mass $\mathcal{L}\supset -\fr 1 2 m_n \bar{n}^c_L n_L$. The dark parity of $\phi'$ is $D_P(\phi'^+)=1$ and $D_P(\phi'^0)=-1$. The $D_P$ conservation implies $\langle \phi'^0\rangle =0$. Hence, the Dirac mass that couples $\bar{\nu}_L n^c_L$ is prevented. Further, the $U(1)_N$ symmetry suppresses the quadratic $\phi^\dagger \phi'$ and quartic $(\phi^\dagger \phi')^2$ terms. However, the potential includes $V\supset \mu \phi^\dagger \phi' \eta^*+\mu' \chi^*\eta^2$, where $\mu,\mu'$ have a mass dimension. When $\chi$ develops a large VEV, say $\La\gg \mu\sim \mu'$, which breaks the dark charge by two unit, the field $\eta$ carries a corresponding large mass $\sim \sqrt{\mu'\La} $ induced by the second term, while the first term produces an effective coupling $\la' (\phi^\dagger \phi')^2$ after integrating $\eta$ out, where $\la'\sim \mu^2/(\mu'\La)\ll 1$ as desirable. Alternative to the seesaw mechanism, the neutrino mass $m_\nu$ that couples $\nu_L \nu_L$ is now generated by an one-loop diagram with the exchange of $\phi'$ and $n_L$, similar to the standard scotogenic theory.\footnote{Including the seesaw contribution, the neutrino mass is given by a combination of type I and II seesaw mechanism.} Last, but not least, the term $\phi^\dagger \phi'$ is always discarded by the $D_P$ conservation, while $n_L$ and $\phi'$ are light, responsible for dark matter.\footnote{A similar realization of scotogenic dark matter in the other context with residual matter parity recently interpreted in \cite{svs}.} The freeze-in mechanism for $n_L$ abundance now works with the decay $\tilde{\phi}'$ to $l n$. 
\item Minimal dark matter \cite{mdm}: The dark matter candidate is contained in a scalar or fermion multiplet that does not singly couple to standard model particles at the renormalizable level, analogous to the proton case. This requires a large representation dimension ($\geq 5$) for the multiplet. However, with the flipped extension, a scalar or fermion triplet can obey a minimal dark matter. Take a scalar triplet with $Y=0$ and $N=1$, \be \Delta = 
\left(\begin{array}{cc} 
\Delta^+_1 & \fr{1}{\sqrt{2}}\Delta^0_2 \\
\fr{1}{\sqrt{2}} \Delta^0_2 & \Delta^-_3 
\end{array}\right).\ee Thus, $\Delta$ does not couple to the standard model Higgs doublet $\phi \phi^*$ due to the $U(1)_N$ conservation. Additionally, the difference in weak isospin leads to $D_P(\Delta_{1,3})=1$, while $D_P(\Delta_2)=-1$. Hence, $\Delta_2$ cannot develop a VEV due to the $D_P$ conservation, hereby suppressing a potential decay mode coming from breaking, $\Delta_2\rightarrow HH$, where $H$ is the Higgs boson. The field $\Delta_2$ is stable, responsible for dark matter, given that it is the lightest component in $\Delta$ due to quantum corrections \cite{mdm}. Since $\Delta_2$ which has $T_3=Y=0$ does not interact with $Z$, the large direct detection cross-section due to $Z$ exchange is suppressed. Note that in this minimal dark matter setup, we need not require $\Delta_2$ to be the lightest odd particle. Hence, the triplet candidate can be heavy enough, responsible for thermal dark matter. Of course, the neutrino mass comes from the above seesaw, while the fields $n,\eta$ are not necessarily introduced to this model.                       
\een                

To conclude, flipping a symmetry is necessarily determined by an extra noncommutative charge. This principle implies a dark charge, besides usual electric charge. The presence of dark charge leads to right-handed neutrinos, charged under dark charge, and they gain large Majorana masses due to dark charge breaking, not $B-L$ breaking. This produces small neutrino masses and resultant dark parity, implying a dark matter candidate lighter than electron. The flipped symmetry breaking scenario at large scale generates asymmetric dark matter with mass at $0.1$ MeV, besides baryon asymmetry, both by standard leptogenesis. Whereas, the flipped symmetry breaking scenario at TeV scale recognizes freeze-in dark matter with mass at $0.1$ MeV, governed by darkodynamics, $U(1)_N$. The flipping principle significantly improves the theoretical and phenomenological aspects of the scotogenic and minimal dark matter. The implication of multicomponent dark matter and a solution dedicated to the doublet-triplet splitting problem in GUT due to noncommutative dark parity are worth exploring.                 

This research is funded by Vietnam National Foundation for Science and Technology Development (NAFOSTED) under grant number 103.01-2019.353.

\end{document}